\documentclass[twocolumn,groupedaddress,aps,prx,superscriptaddress,amssymb,amsmath,amsfonts,10pt]{revtex4-2}

\usepackage{bm}
\usepackage{graphicx}
\usepackage{subfigure}
\usepackage{algorithm}
\usepackage{algpseudocode}
\usepackage{multirow}
\usepackage{wasysym} 

\begin{document}

\title{ Reconcile the Bulk Metallic and Surface Insulating\\ State in 1\emph{T}-TaSe$_2$}

\author{Wenhao Zhang}
\affiliation{Zhejiang Province Key Laboratory of Quantum Technology and Device, Department of Physics, Zhejiang University, Hangzhou 310027, China}

\author{Zongxiu Wu}
\affiliation{Zhejiang Province Key Laboratory of Quantum Technology and Device, Department of Physics, Zhejiang University, Hangzhou 310027, China}
\author{Kunliang Bu}
\affiliation{Zhejiang Province Key Laboratory of Quantum Technology and Device, Department of Physics, Zhejiang University, Hangzhou 310027, China}

\author{Ying Fei}
\affiliation{Zhejiang Province Key Laboratory of Quantum Technology and Device, Department of Physics, Zhejiang University, Hangzhou 310027, China}
\author{Yuan Zheng}
\affiliation{Zhejiang Province Key Laboratory of Quantum Technology and Device, Department of Physics, Zhejiang University, Hangzhou 310027, China}
\author{Jingjing Gao}
\affiliation{Key Laboratory of Materials Physics, Institute of Solid State Physics, Chinese Academy of Sciences, Hefei 230031, China}
\affiliation{University of Science and Technology of China, Hefei 230026, China}
\author{Xuan Luo}
\affiliation{Key Laboratory of Materials Physics, Institute of Solid State Physics, Chinese Academy of Sciences, Hefei 230031, China}
\author{Zheng Liu}
\affiliation{Institute for Advanced Study, Tsinghua University, Beijing 100084, China}
\author{Yu-Ping Sun}
\affiliation{Key Laboratory of Materials Physics, Institute of Solid State Physics, Chinese Academy of Sciences, Hefei 230031, China}
\affiliation{High Magnetic Field Laboratory, Chinese Academy of Sciences, Hefei 230031, China}
\affiliation{Collaborative Innovation Center of Advanced Microstructures, Nanjing University, Nanjing 210093, China}
\author{Yi Yin}
\email{yiyin@zju.edu.cn}
\affiliation{Zhejiang Province Key Laboratory of Quantum Technology and Device, Department of Physics, Zhejiang University, Hangzhou 310027, China}
\affiliation{Collaborative Innovation Center of Advanced Microstructures, Nanjing University, Nanjing 210093, China}

\begin{abstract}
The origin of different electronic states of 1\emph{T}-TaS$_2$ and 
1\emph{T}-TaSe$_2$ remains controversial due to the complicated correlated 
electronic properties. We apply scanning tunneling microscopy to study the 
electronic state of bulk 1\emph{T}-TaSe$_2$. Both insulating and metallic 
states are identified in different areas of the same sample.
The insulating state is similar to that in 1\emph{T}-TaS$_2$, concerning 
both the $\it{dI/dV}$ spectrum and the orbital texture. With detailed 
investigations in single-step areas, the electronic state measured on 
the upper-layer surface is found to be associated with differnt stacking 
orders and the lower layer's electronic state. 
The insulating state is most possibly a single-layer property, 
perturbed to a metallic state by particular stacking orders. Both the 
metallic and large-gap insulating spectra, 
together with their corresponding stacking orders, are stable states 
in 1$\emph{T}$-TaSe$_2$. The connected metallic 
areas lead to the metallic transport behavior. We then reconcile the 
bulk metallic and surface insulating 
state in 1\emph{T}-TaSe$_2$. The rich phenomenon in 1\emph{T}-TaSe$_2$ 
deepens our understanding of the correlated 
electronic state in bulk 1\emph{T}-TaSe$_2$ and 1\emph{T}-TaS$_2$.
\end{abstract}

\maketitle

\section{Introduction}

Layered transition metal dichalcogenides (TMDs) 1\emph{T}-TaS$_2$ 
and 1\emph{T}-TaSe$_2$ are two widely-studied charge 
density wave (CDW) materials in past few 
decades~\cite{CDW, doping_CDW, electronic_pro, CDW_origin, Mott_STM, STM_Tem_dep, Mott_ARPES, MottToSC, combine_TaS, Disorder_Mott}.
Both materials are at a commensurate CDW (CCDW) phase with 
a $\sqrt{13}\times\sqrt{13}$ reconstruction at low temperature.
A single unpaired electron exists in each star of David (SD), 
the basic element of the CCDW super-lattice, resulting in 
a half-filled system. The electron-electron 
correlation can split the half-filled band to the lower and 
upper Hubbard band (LHB and UHB) with an insulating gap, 
called the Mott insulator~\cite{MIT}. In the bulk 1\emph{T}-TaS$_2$, 
different experimental techniques confirm the insulating gap
around the Fermi level, consistent with the transport result 
in 1\emph{T}-TaS$_2$~\cite{Mott_STM,STM_Tem_dep, Mott_ARPES}.
The insulating gap nature of 1\emph{T}-TaS$_2$ is 
still under debate~\cite{orbital_texture,stacking,lee2019origin,
Mott_DFT_PRL,Mosica_state,NanoScale,butler2020mottness,double_PRX,Yeom_PRL,bandinsulator}, 
and is comprehensively studied in 
our other work~\cite{zxwu21}. 

For the bulk 1\emph{T}-TaSe$_2$, electronic state measured with different
techniques are more 
complicated~\cite{ ARPES_3d_band, STM_site_metallic, surface_Mott_ARPES, surface_Mott_STM}.
Both scanning tunneling microscopy (STM) and angle-resolved
photoemission spectroscopy (ARPES) have detected an insulating state on the exposed
surface~\cite{surface_Mott_ARPES, surface_Mott_STM}.
However, the resistivity measurement shows a metallic behavior in the CCDW phase,
also referred to as a bulk metallic state. 
The bulk metallic state and the surface insulating state have to be reconciled
in the same material, which is still far from a complete understanding.
On the other hand, Se atoms in 1\emph{T}-TaSe$_2$
share the same valence as S atoms in 1\emph{T}-TaS$_2$~\cite{TaS_DFT, TaSe_DFT}.
Whether the surface insulating state in 1\emph{T}-TaSe$_2$ is similar to 
the correlation-induced state in 1\emph{T}-TaS$_2$ needs to be clarified.
The insulating state in the single-layer film of
1\emph{T}-TaSe$_2$/TaS$_2$, grown by molecular beam epitaxy (MBE), makes the
problem more complicated and fascinating~\cite{TaSe_Chenyi,TaSe_jisuaihua}.

Here we perform a detailed STM experiment to explore the electronic state of
pristine 1\emph{T}-TaSe$_2$ at low temperature. We find both metallic and insulating
states frequently on the exposed surface of 1\emph{T}-TaSe$_2$. The insulating spectrum in
1\emph{T}-TaSe$_2$ is very similar to that in 1\emph{T}-TaS$_2$, from 
the $\it{dI/dV}$ spectrum to the spatial orbital texture. 
With multi-step cleaved surfaces, we measure and obtain three different 
types of $\it{dI/dV}$ spectra, including the large-gap, small-gap, 
and metallic spectra.
The discrepancy of different electronic states in 1\emph{T}-TaSe$_2$ is found to be
associated with different stacking orders. 

To infer the intrinsic stacking-order results in the bulk material, we focus on the flat 
single-step area and extend the measurement to areas away from the step edge. Both the 
large-gap insulating state and the metallic state are found to be a bulk property, while 
the small-gap spectrum only exists within a small area around the step edge. 
When the lower layer is metallic or insulating, we conclude the correspondence between 
the stacking order and the upper-layer electronic state, and list
the number of occurrences of different situations. With the large-gap insulating state 
most possibly the single-layer property of 1\emph{T}-TaSe$_2$, the metallic state could originate
from the stacking-order-induced interlayer coupling and perturbation to the insulating state. 
The frequent appearance of metallic state in 1$\emph{T}$-TaSe$_2$ leads to the macroscopic metallic transport results.
The main difference between 1$\emph{T}$-TaS$_2$ and 1$\emph{T}$-TaSe$_2$ is that this perturbed 
metallic state is energetically stable in 1$\emph{T}$-TaSe$_2$ but not stable in 1$\emph{T}$-TaS$_2$. 
It requires a further theoretical study to explore how the stacking order induces this 
perturbation and the different stability of this metallic state in two materials.

\begin{figure}
    \centering
    \includegraphics[width=1.0\columnwidth]{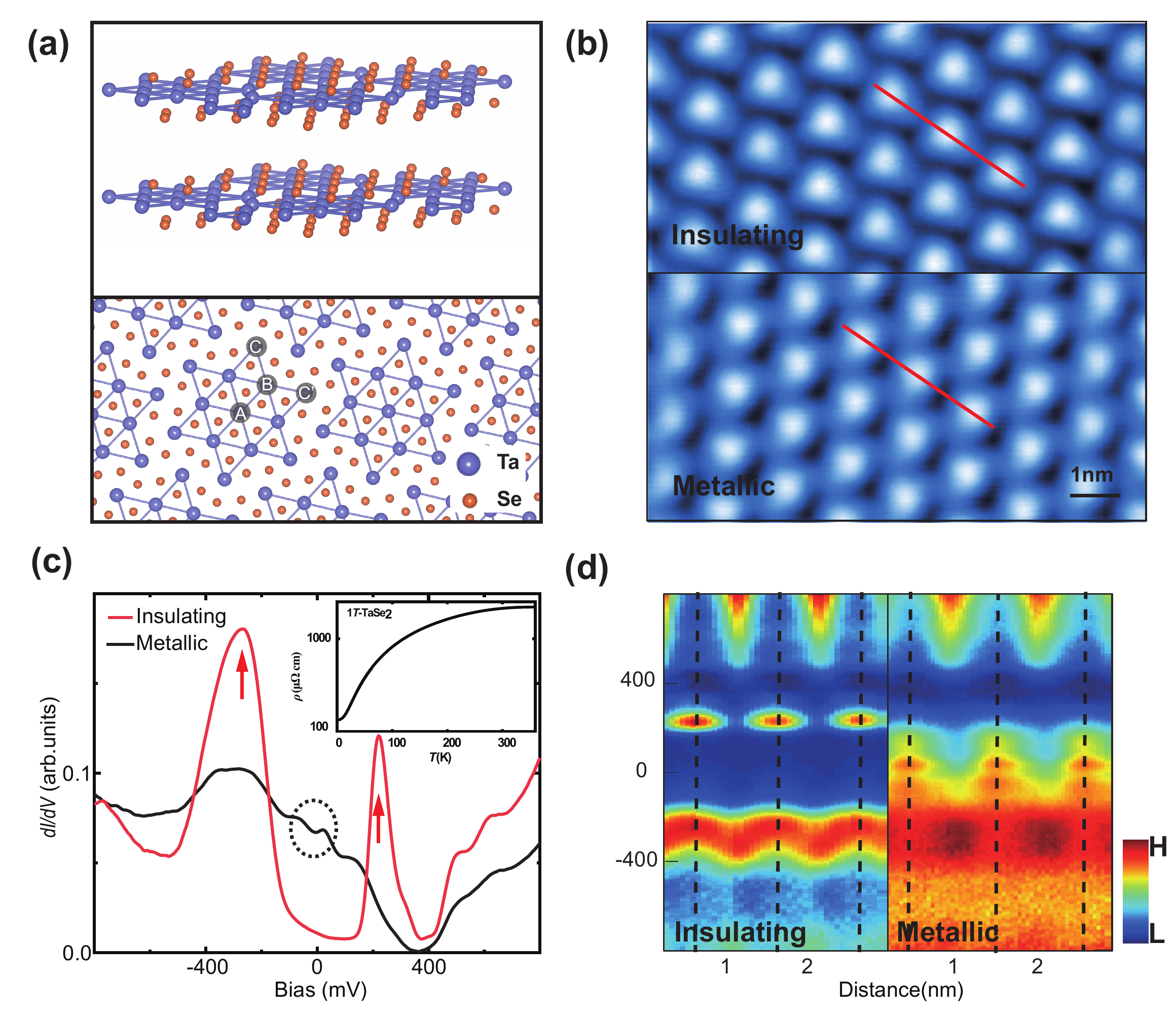}
    \caption{Overview of structural and electronic properties of 1$\emph{T}$-TaSe$_2$.
    (a) Crystal structure of 1$\emph{T}$-TaSe$_2$. The side-view sketch is
    shown in the upper panel. The star-of-David (SD) is formed by 13 Ta atoms in the CCDW phase, with the top-view sketch shown in the lower panel. Different Ta atoms are labeled by A, B, C, and C'.
    (b) Two $10\times5$ nm$^2$ topographies taken in the area of
    insulating (upper panel) and metallic (lower panel) state.
    (c) Typical $\it{dI/dV}$ spectra taken in the area of insulating and metallic state
    (setup condition: $V_\mathrm{b}=-600$ mV, $I_\mathrm{s}=500$ pA). The inset shows resistivity of bulk 1$\emph{T}$-TaSe$_2$ at the range of $2-360$ K, exhibiting a metallic behavior at low temperature.
    (d) Two linecut maps crossing three SDs along the red lines in (b). The periodic $\it{dI/dV}$ evolution
    is displayed on the horizontal direction, in the bias voltage range
    of $[-800, 800]$ mV.
    }\label{fig_01}
\end{figure}

\section{Experiment and Results}

High-quality single crystal 1$\emph{T}$-TaSe$_2$ was grown with the iodine transport agent.
In the STM experiment, 1$\emph{T}$-TaSe$_2$ were cleaved in situ in an ultra-high vacuum chamber
(base pressure $4 \times 10^{-11}$ mBar) at liquid nitrogen temperature and transferred
to the STM head immediately. Topographies in this paper were collected with a bias
voltage $-600$ mV and setpoint current $500$ pA at liquid nitrogen temperature
($\sim{77}$ K). The differential conductance $\it{dI/dV}$ spectra were acquired by the
standard lock-in technique, with a voltage modulation of $10$ mV at $1213.7$ Hz frequency.

Bulk single crystal 1$\emph{T}$-TaSe$_2$ has a CdI$_2$-type layered structure
[upper panel of Fig.~\ref{fig_01}(a)]. In each unit structure,
the Ta atomic layer is sandwiched between two Se atomic layers, and the Ta atoms form a
triangular lattice in the plane. In the CCDW phase, every $13$ Ta atoms are
regrouped into a cluster called star-of-David (SD), and the SDs form an enlarged triangular
super-lattice [lower panel in Fig.~\ref{fig_01}(a)]. 
The center and nearest Ta atoms are labeled by A and B respectively, 
while the inequivalent next nearest Ta atoms are labeled by C, and C'. 
The electronic state is strongly
modified by the CCDW phase. The bands from Ta atoms are folded into three sub-bands with
two filled and one half-filled~\cite{ARPES_3d_band}. In the bulk 1$\emph{T}$-TaS$_2$,
the half-filled band is split into the LHB and UHB
with a Mott insulating gap, consistent with the insulating state in the resistivity
measurement. In contrast, the resistivity measurement
of 1$\emph{T}$-TaSe$_2$ [Fig.~\ref{fig_01}(c) inset]
shows a metallic state at low temperature~\cite{CDW, doping_CDW, electronic_pro}.
The discrepancy of resistivity results of 1$\emph{T}$-TaS$_2$ and 1$\emph{T}$-TaSe$_2$ has been
explained by the different $p$-derived valence bands of S and
Se~\cite{pseudogap_aebi, hybrid_ARPES, TaS_Se_dope, SC_TaSSe, QiaoPRX17}.
However, previous ARPES and STM experiment of bulk
1$\emph{T}$-TaSe$_2$ also observed that a surface metallic to Mott insulating phase transition
occurs when the temperature is reduced to 260 K~\cite{surface_Mott_ARPES, surface_Mott_STM}.
The reason for the coexistence of bulk metallic and surface insulating state remains
unresolved in 1$\emph{T}$-TaSe$_2$.


In our STM experiment, the measurement temperature ($\approx 77$ K) is far below the
CCDW transition temperature. Insulating and metallic $\it{dI/dV}$ spectra
both exist in different areas of each sample we measured. Figure~\ref{fig_01}(b) are
two representative topographies with different electronic states. They share
the same triangular SD super-lattice array, with each bright spot corresponding to one SD.
Figure~\ref{fig_01}(c) shows the typical insulating and
metallic spectra measured at the center of one SD. 
The insulating $\it{dI/dV}$ spectrum
is very similar to the Mott insulating spectrum measured in bulk 1$\emph{T}$-TaS$_2$.
The LHB and UHB peaks are located at $-300$ and $230$ mV.
The gap size is around $530$ mV, about 100 mV larger than that of
bulk 1$\emph{T}$-TaS$_2$~\cite{QiaoPRX17,eeinteraction_TaS, bucp18}.
The CDW valley roughly at $400$ mV shows an approximate zero value, suggesting a strong
CDW modulation in 1$\emph{T}$-TaSe$_2$~\cite{TaSe_DFT}. 
The metallic $\it{dI/dV}$ spectrum displays a rough
linear decay of density-of-state (DOS) around the Fermi energy (zero bias). 
A weak dip feature
appears around the zero bias. The deep valley at $400$ mV coincides with the
CDW valley in the insulating $\it{dI/dV}$ spectrum.

To show the spatial homogeneity, we present the linecut maps for both
the insulating and metallic states. In Fig.~\ref{fig_01}(d), the horizontal axis is the
position along the two red lines crossing three SDs in Fig.~\ref{fig_01}(b), and the vertical
axis is the STM bias ranging from $-800$ to $800$ mV. For the insulating spectra,
the LHB and UHB peaks are evident at SD centers, resulting in a periodic intensity
variation along the line. The CDW feature above 400 mV also shows an apparent periodic
variation, but with maximum peaks shifting by half the period. This phenomenon is
similar to that in 1$\emph{T}$-TaS$_2$~\cite{QiaoPRX17, eeinteraction_TaS}.
The Hubbard bands originate from the central Ta
orbital, with LHB and UHB peaks showing maximum values at SD centers.
The CDW bands are from the surrounding Ta orbital, with CDW peaks showing maximum values
at SD rims. In the linecut map of metallic spectra, the zero-bias dip
also shows a periodic variation, and the maximum values are located at
SD centers. We expect that the two peaks
of the zero-bias dip in the metallic
spectrum are also related to the central Ta orbital.

\begin{figure}
    \centering
    \includegraphics[width=1\columnwidth]{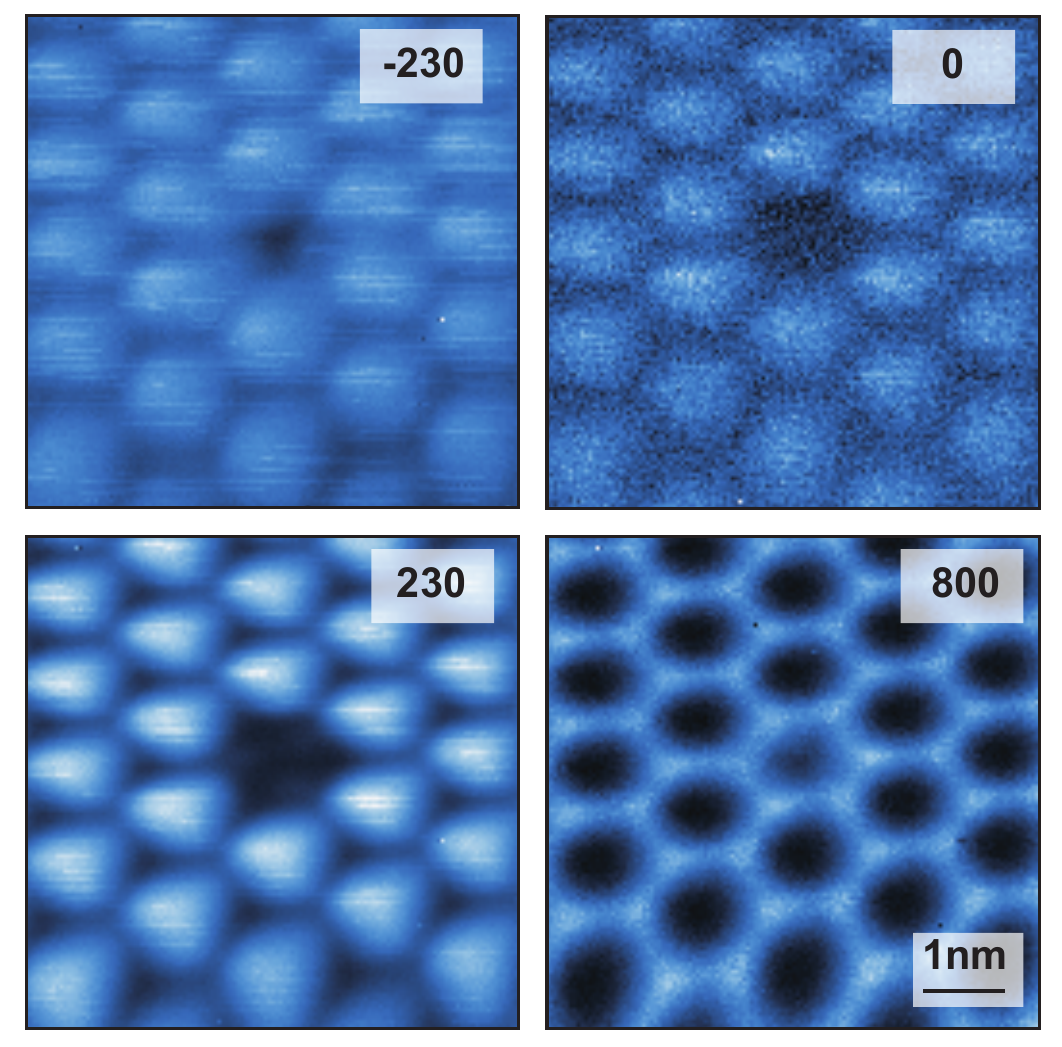}
    \caption{The $\it{dI/dV}$ maps at several selected energies ($-230, 0, 230,800$ mV) for a $6\times 6$ nm$^2$ area with insulating state. A defective SD exists in the middle of this area.
    }
    \label{fig_02}
\end{figure}


Figure~\ref{fig_02} displays the differential conductance $\it{dI/dV}$ maps at
several selected energies for a $6\times 6$ nm$^2$ area with insulating state.
A special point defect exists at one SD in the middle of the frame, which 
is identified to a central Ta vacancy in 1$\emph{T}$-TaS$_2$~\cite{zxwu21}.
The $\it{dI/dV}$ maps reveal more
spatial characteristics than the linecut map crossing SDs.
With the bias voltage at $-230$ mV and $230$ mV, these two $\it{dI/dV}$ maps are close to
the LHB and UHB, respectively. Both maps display a triangular lattice pattern with the bright spots positioned at SD centers, consistent with that the Hubbard band mainly originates from the central Ta orbital. 
When moving to the lower energy or higher energy,
the $\it{dI/dV}$ maps reveal a honeycomb orbital texture (e.g. in the map at $800$ mV),
consistent with that the CDW bands are from the surrounding Ta orbitals. The
energy-dependent orbital texture for the insulating state is similar to that in the
bulk 1$\emph{T}$-TaS$_2$~\cite{QiaoPRX17}. The single-layer MBE-grown
1$\emph{T}$-TaSe$_2$ shows a slightly different energy-dependent orbital
texture despite of the similar insulating spectrum~\cite{TaSe_Chenyi,TaSe_jisuaihua}.
The similar insulating spectra in the single-layer and bulk samples suggest
that the insulating state is most likely an intrinsic single-layer property.

\begin{figure}
    \centering
    \includegraphics[width=1.0\columnwidth]{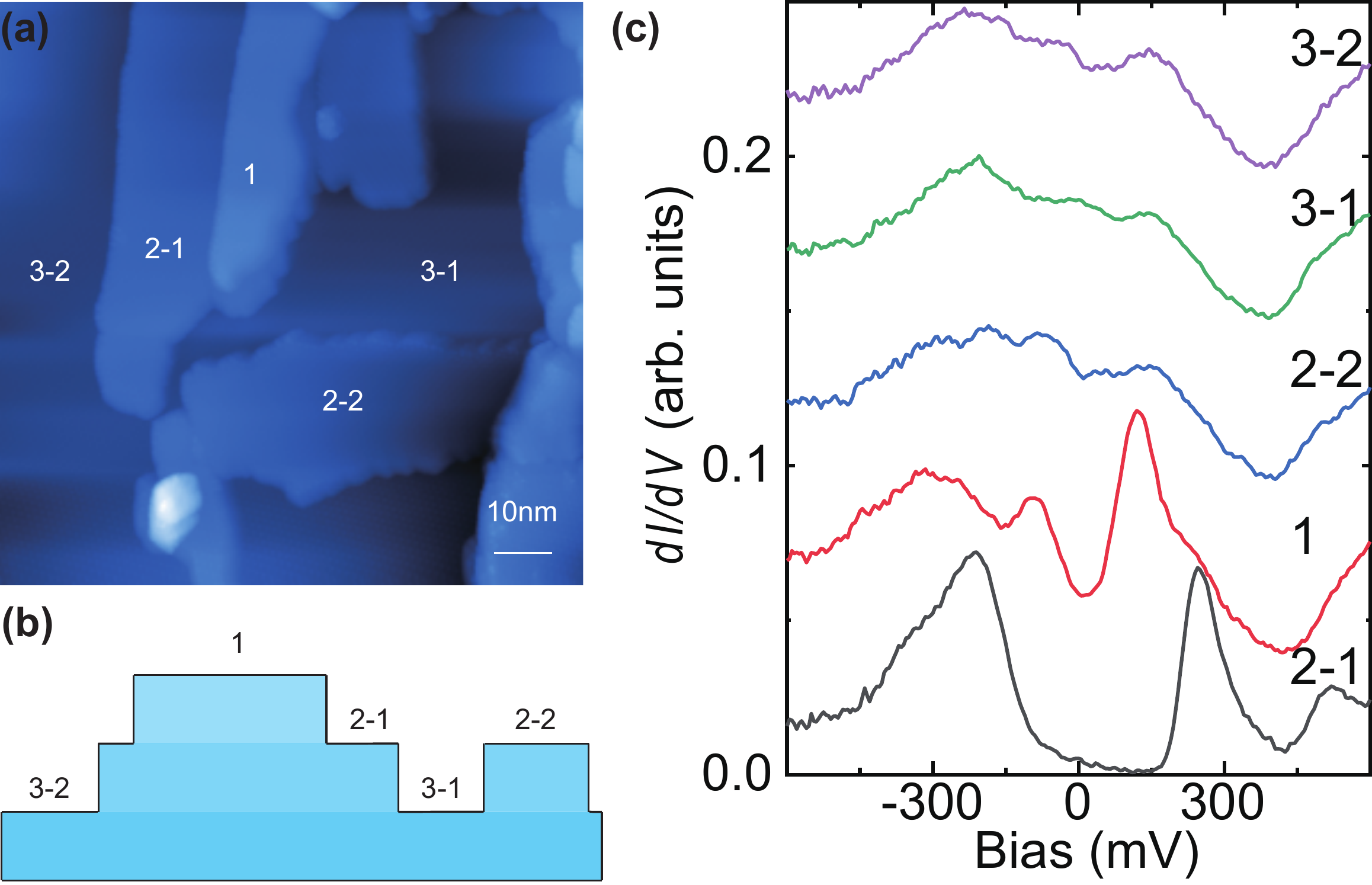}
    \caption{
        Topography and corresponding electronic spectra in a multi-step area.
        (a) A $100\times 100$ nm$^2$ topography, with a schematic side-view shown in (b).
        (c) Typical  $\it{dI/dV}$ spectra on different regions in (a). The spectrum in region 3-1, 3-2 and 2-2
        shows a normal metallic state of 1\emph{T}-TaSe$_2$, while the spectra in 
region 2-1 and 1 show a large-gap state and a small-gap state, respectively.
    }
    \label{fig_03}
\end{figure}


The reason for the different electronic states in different
regions is still a mystery.
Stacking order has been taken into account in some recent research
works~\cite{orbital_texture,stacking,lee2019origin,Mott_DFT_PRL,Mosica_state,NanoScale, butler2020mottness,double_PRX,Yeom_PRL,bandinsulator}.
In our other work~\cite{zxwu21}, stacking order has been detailed studied in 1$\emph{T}$-TaS$_2$.
We apply a similar procedure to study the step areas on cleaved bulk 1$\emph{T}$-TaSe$_2$~\cite{zxwu21}.
Figure~\ref{fig_03}(a) shows a special large area topography with multi-step environment,
taken on the exposed surface of 1\emph{T}-TaSe$_2$. 
A schematic side view of
this multi-step area is shown in Fig.~\ref{fig_03}(b), which is derived from the
height profiles in topography. As shown in Fig.~\ref{fig_03}(c), the $\it{dI/dV}$
spectra measured on different step planes exhibits different electronic states.

For the area in Fig.~\ref{fig_03}(a), the $\it{dI/dV}$ spectrum on the bottom 
layer (region 3-1 and region 3-2) is a metallic one similar to that in Fig.~\ref{fig_01}(c). 
On top of this layer, the $\it{dI/dV}$ spectra exhibit different electronic state, 
insulating in region 2-1, and metallic in region 2-2.
Different electronic states in the same plane (region 2-1 and 2-2)
imply different stacking orders exist between the same two layers. 
Furthermore, the $\it{dI/dV}$ spectrum on region 1 has a reduced
gap with peaks at $-70$ and $140$ mV. Although the zero-bias
conductance of this type of spectrum is enhanced to a significant
value, the curve shape is consistent with the small gap spectrum in 1$\emph{T}$-TaS$_2$~\cite{zxwu21}. 
This phenomenon reminds us the similar three types of
spectra in the stacking order study in 1$\emph{T}$-TaS$_2$~\cite{zxwu21}.

\begin{figure}
    \centering
    \includegraphics[width=1\columnwidth]{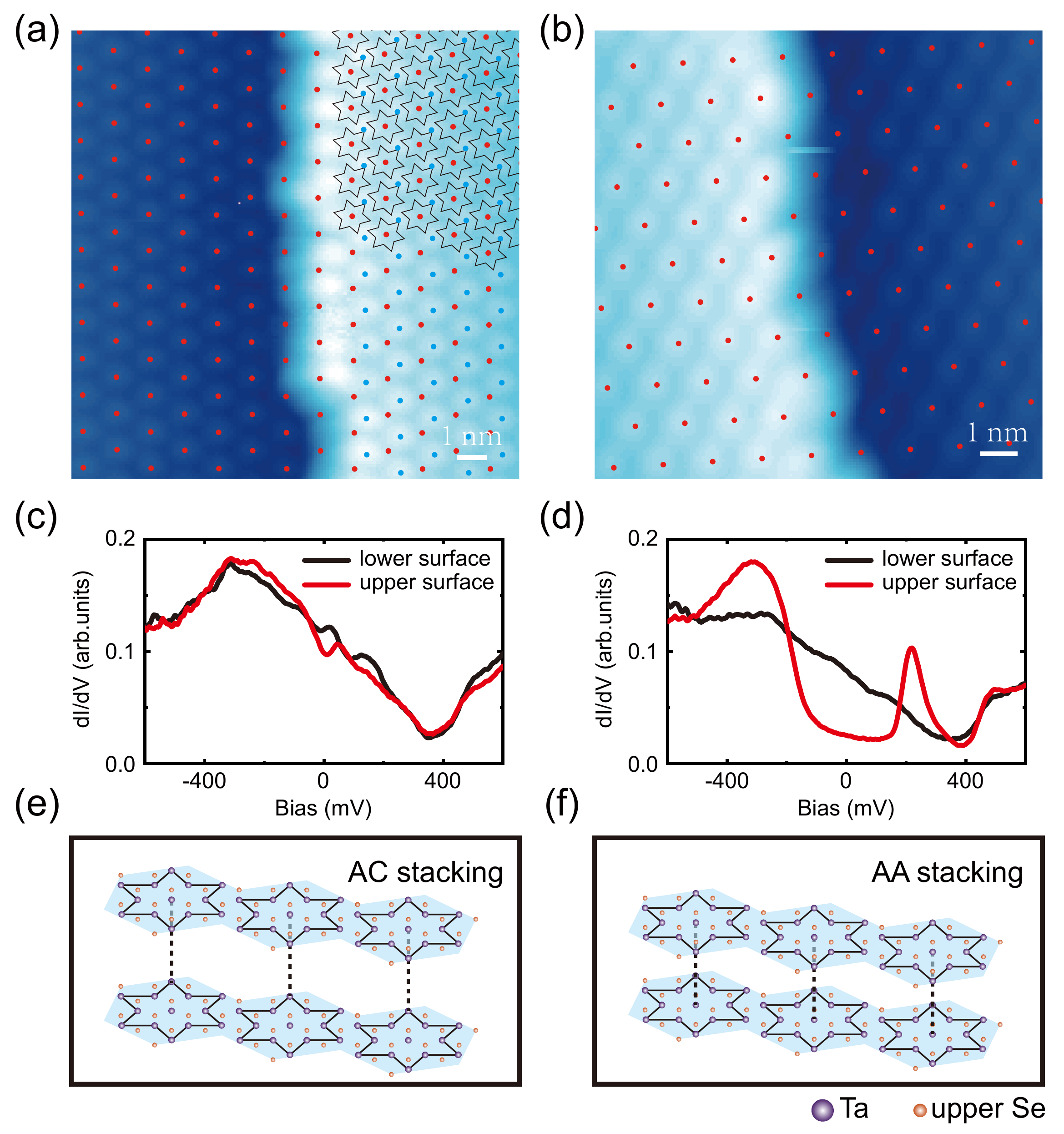}
    \caption{
        The stacking orders and corresponding $\it{dI/dV}$ spectra with a lower-layer metallic spectrum.
        (a) and (b) STM topographies of a single-step area. The red (blue) dots mark SD centers in the lower (upper) layer. The SD centers in the upper layer are aligned with
        the SD outer-corner sites (AC-stacking order), or center sites (AA-stacking order) in the lower layer. 
        (c) and (d) The red and black
        lines are the $\it{dI/dV}$ spectra on the upper and lower layers in (a) and (b). 
        (e) and (f) The schematic diagrams of the stacking orders in (a) and (b).
    }
    \label{fig_04}
\end{figure}


To study the stacking order, we focus on the flat single-step area which
renders a determination of the relative displacement of SD super-lattices in the
upper and lower layers. The method is similar to that in our other work~\cite{zxwu21}.
The single-step area also enables us to extend the measurement away from the step edge to 
check whether the result is more related to the bulk property. In multiple experiments, we 
only occasionally observe the small-gap spectrum, which also exsits within 
a small area around the step edge or in a small island area [Fig.~\ref{fig_03}(a)]. Then
we mainly concentrate attention on the large-gap and metallic spectra, which are more
related to the intrinsic property of the bulk material. For the following shown 
examples, they are selected from multiple measurements on different samples. 

We first consider the situation with a metallic spectrum on the lower layer.
Figure~\ref{fig_04}(a) and \ref{fig_04}(b) show two single-step area topographies, which
include triangular SD super-lattices in both the lower
layer and the bright upper layer. The spectra measured
on the lower and upper layers are shown in Fig.~\ref{fig_04}(c) and \ref{fig_04}(d).
On the upper layer, the electronic state shows a metallic
spectrum in Fig.~\ref{fig_04}(c). The metallic $dI/dV$ spectra of the two layers 
are basically the same with slight difference near the Fermi level.
The corresponding stacking order is the AC-stacking, which means that
the SD center of the upper layer is vertically aligned with the
outer-corner site of SD in the underlying layer as shown in Fig.~\ref{fig_04}(e). 
The metallic-metallic steps are frequently observed during our experiments.
Another commonly observed case is the insulating upper layer 
observed with AA-stacking [Fig.\ref{fig_04}(d) and \ref{fig_04}(f)]. 
It is worth to mention that although the small-gap spectrum 
is rarely observed above the metallic layer, 
the corresponding stacking order is the AB-stacking.
Up to our current experiments, the surface electronic states 
above a metallic layer all shows a one-to-one correspondence with stacking orders.




\begin{figure*}
    \centering
    \includegraphics[width=1.38\columnwidth]{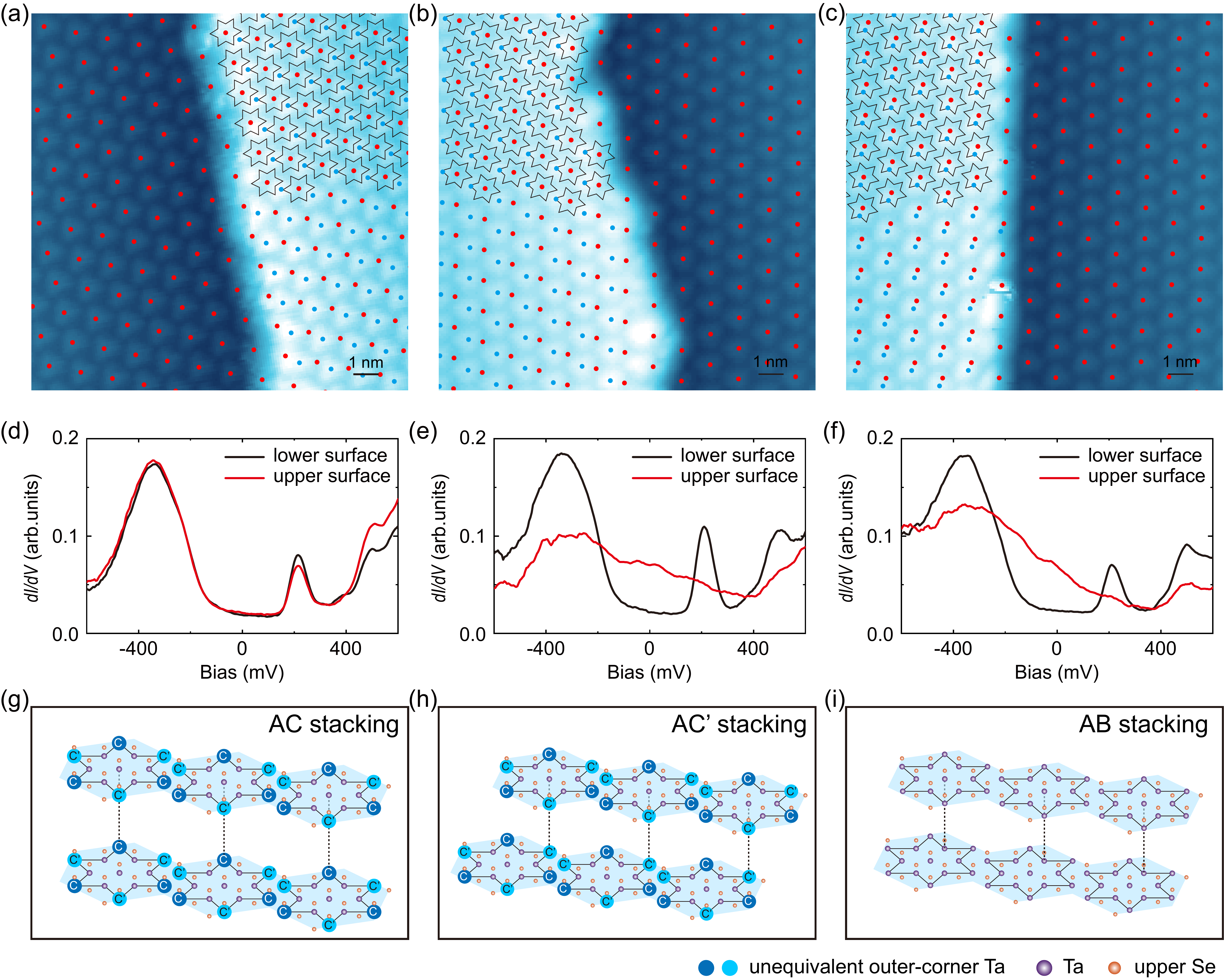}
    \caption{
        The stacking orders and corresponding $\it{dI/dV}$ spectra with a lower-layer insulating spectrum.
        (a) STM topography of a single-step area. The red (blue) dots mark SD centers
        in the lower (upper) layer. The SD centers in the upper layer are aligned with
        the SD outer-corner sites in the lower layer, or in the AC-stacking order. (b), (c) 
        STM topographies of two other single-step areas, in the AC'-stacking or AB-stacking orders
        as shown in the superimposed SD centers. (d)-(f) The red and black
        lines are the $\it{dI/dV}$ spectra on the upper and lower layers in (a)-(c). (h)-(i) The
        schematic diagrams of the stacking orders in (a)-(c).
    }
    \label{fig_05}
\end{figure*}

We further consider the situation with an insulating spectrum on the lower layer.
The insulating-insulating steps, which are most frequently observed 
in 1$\emph{T}$-TaS$_2$ are also commonly seen in 1$\emph{T}$-TaSe$_2$.
Different from 1$\emph{T}$-TaS$_2$, the AA-stacking of insulating-insulating 
step only occasionally appears in 1$\emph{T}$-TaSe$_2$. 
The dominate stacking order is the AC-stacking [Fig.~\ref{fig_05}(a), \ref{fig_05}(d), and \ref{fig_05}(g)].
Figure~\ref{fig_05}(b) and \ref{fig_05}(c)
show two single-step area topographies, with the metallic spectra
measured on the upper layers [Fig.~\ref{fig_05}(e) and \ref{fig_05}(f)].
After carefully determining the relative shift of SD super-lattices, the 
metallic layer in Fig.~\ref{fig_05}(b) corresponds to the AC'-stacking, 
which is unequivalent with the AC-stacking in Fig.~\ref{fig_05}(a). 
Figure~\ref{fig_05}(c) corresponds to the AB-stacking order. 
With multiple measurements, we find that these three
cases are all frequent and constantly reproducible.

\begin{table*}[]
    \centering
    \begin{tabular}{|c|c|c|c|}
    \hline
    \textbf{\begin{tabular}[c]{@{}c@{}}lower-layer \\electronic state\end{tabular}} & \textbf{\begin{tabular}[c]{@{}c@{}}upper-layer\\ electronic state\end{tabular}} & \textbf{stacking order} & \textbf{number} \\ \hline
    \multirow{2}{*}{metallic} & metallic & AC & 6 \\ \cline{2-4} 
     & large-gap insulating & AA & 4 \\ \hline
    \multirow{8}{*}{large-gap insulating} 
     & \multirow{3}{*}{large-gap insulating} 
     & AC & 7 \\ \cline{3-4} 
     &  & AA & 1 \\ \cline{2-4} 
     & \multirow{3}{*}{metallic} 
     & AC' & 5 \\ \cline{3-4} 
     &  & AB & 5 \\ \hline
    \end{tabular}%
    \caption{Summary of different stacking orders observed during the experiments}
    \label{tab_01}
\end{table*}

All different stacking orders and corresponding electronic state 
of upper and lower layers are summarized in Table~\ref{tab_01}. 
As the table shows, the stacking order shows one-to-one correspondence 
with the upper-layer electronic state, when the lower layer is metallic. 
When the lower layer is insulating, two different stacking orders 
can lead to the same upper-layer electronic state. 
A similar case has happened in 1$\emph{T}$-TaS$_2$~\cite{zxwu21}, 
in which both AA and AC-stacking lead to the 
insulating-insulating step. Considering the single-layer 
property of the large-gap spectrum, both AA and AC-stacking 
may lead to a negligible interlayer coupling. In contrast, 
the AC' and AC-stacking result in a similar perturbed and 
metallic upper-layer spectrum. The upper-layer spectrum is not 
only related to the stacking order, but also to the spectrum
of the lower layer. The AC-stacking order could correspond to 
a metalic spectrum in Fig.~\ref{fig_04}(a) but an insulating 
spectrum in Fig.~\ref{fig_05}(a).  

There are long-standing puzzles about the insulating state in bulk 1$\emph{T}$-TaS$_2$ and 1$\emph{T}$-TaSe$_2$. According to 
the first-principle calculation, there is a conducting band along the $\Gamma A$ direction, which leads to a metallic state near the Fermi level. 
However, the large insulating gap near the Fermi level has been identified by different experimental methods.
Recently, some experiments reveal possible double-layer stacking in 1$\emph{T}$-TaS$_2$, which conjures to a band insulator scenario~\cite{butler2020mottness,double_PRX,Yeom_PRL}.
In our comprehensive results, the AC-stacking in Fig~\ref{fig_05}(a) corresponds to two large-gap spectra on both the upper 
and lower terraces, which dispels that the large gap is from the unit-cell 
doubling of two AA-stacked layers~\cite{butler2020mottness,double_PRX,Yeom_PRL}.
In fact, different stacking orders appear randomly in our samples. The upper-layer electronic state is 
determined by the stacking orders and the electronic state of adjacent underlying layer. 
The large-gap spectrum is most possibly the intrinsic property of a single-layer 1$\emph{T}$-TaSe$_2$, 
similar to that of MBE grown film of 1$\emph{T}$-TaSe$_2$. The metallic spectrum is 
not intrinsic in the single-layer 1$\emph{T}$-TaSe$_2$, but a modulated spectrum induced 
by particular stacking orders. 

The metallic spectrum in 1\emph{T}-TaSe$_2$ shows a rough linear DOS near the Fermi level, 
different from the V-shaped metallic spectrum in 1\emph{T}-TaS$_2$. 
The difference is possibly related to the different valence band of S and Se. 
The Se atom valence band has a higher energy level than that of S atom, making it 
easier for the band to hybridize with the $\it{d}$ band of Ta atoms. 
On top of the insulating layer, a metallic layer can be induced by the AC'-stacking 
and AB-stacking orders. Above the metallic layer, the AC-stacking leads to the metallic upper-layer state.  
When the electron correlation is disturbed, the strong charge transfer between 
Ta and Se atoms is possibly dominant, leading to the linear DOS decay near the Fermi level.
The difference between Se and S bands may also make the metallic perturbed state more stable in terms of
energy, while the theoretical explaination is beyond our current capability and needs
further investigations.   

In 1$\emph{T}$-TaSe$_2$, both the metallic and large-gap insulating states
can extend all the way at the step edge, although only the large-gap spectrum
can extend all the way in 1$\emph{T}$-TaS$_2$. 
The difference is consistent with that both insulating state and metallic state 
can appear at different regions
in the same sample in 1$\emph{T}$-TaSe$_2$, while only the insulating state
dominates in 1$\emph{T}$-TaS$_2$. Please note that although we can only
measure the spectrum on exposed terraced areas, the collected information
helps to predict the electronic state layer-by-layer when we know the state
of underlying layer and the stacking order.

\section{Conclusion}
Combining all above information, we demonstrate that the stacking order plays a crucial
role in determining the electronic state in bulk 1$\emph{T}$-TaSe$_2$. 
The main difference between bulk 1$\emph{T}$-TaS$_2$ and 1$\emph{T}$-TaSe$_2$ 
is that, the large-gap insulating spectrum dominates in
1$\emph{T}$-TaS$_2$. In contrast, many different stacking orders can induce both 
insulating and metallic states depending on the electronic states of the underlying
layer in 1$\emph{T}$-TaSe$_2$. A large enough ratio of metallic state can always yield 
a metallic state in the macroscopic transport measurement.

In pristine 1$\emph{T}$-TaSe$_2$, the large-gap insulating state mainly reflects the intrinsic
single-layer Mott gap, while some spectral details like the nonzero conductance 
around the Fermi energy and the slope from LHB to zero bias position  
may be related to the inter-layer effect along the $c$-axis.
In general, the inter-layer coupling plays a stronger role in 1$\emph{T}$-TaSe$_2$
than in 1$\emph{T}$-TaS$_2$. The extremely easy modulation of the electronic state
can also make the MBE film slightly different from the insulating state in bulk sample, 
possibly related to the substrate-induced effect or the particular growth condition.
The Mott physics in TMD materials 1$\emph{T}$-TaSe$_2$ and 1$\emph{T}$-TaS$_2$ is
proved to be exceptionally rich and deserves further exploration.

\begin{acknowledgments}

\noindent{This work was supported by the National Key Research and Development
Program of China (Grants No. 2019YFA0308602),
the Key Research and Development Program of Zhejiang Province, China (2021C01002),
and the Fundamental Research Funds for the Central Universities in China.
Z.L. thanks the National Nature Science Foundation of China (NSFC-11774196) and Tsinghua
University Initiative Scientific Research Program.
J.J.G., X.L., and Y.P.S. thank the support of the National Key Research and Development
Program of China (Grants No. 2016YFA0300404), the National Nature Science Foundation of
China (NSFC-11674326 and NSFC-11874357), the Joint Funds of the National Natural
Science Foundation of China, and the Chinese Academy of Sciences' Large-Scale Scientific
Facility (U1832141, U1932217, U2032215).}

\end{acknowledgments}

\end{document}